\documentclass[conference]{IEEEtran}
\IEEEoverridecommandlockouts
\usepackage{cite}
\usepackage{amsmath,amssymb,amsfonts}

\usepackage{graphicx}
\usepackage{textcomp}
\usepackage{xcolor}
\def\BibTeX{{\rm B\kern-.05em{\sc i\kern-.025em b}\kern-.08em
    T\kern-.1667em\lower.7ex\hbox{E}\kern-.125emX}}

\usepackage{listings}
\usepackage{algorithm}
\usepackage{hhline}
\usepackage{colortbl}
\usepackage{centernot}
\usepackage{pbox}
\usepackage{amsmath}
\usepackage{amsfonts}
\usepackage{url}
\usepackage{bm}
\usepackage{comment}
\usepackage{graphicx}
\usepackage{subfigure}
\usepackage{epstopdf}
\usepackage{multirow}
\usepackage{color}
\usepackage{tabu}
\usepackage{mdframed}
\usepackage{syntax}
\usepackage{fancyvrb}
\usepackage{multicol}
\usepackage{booktabs,dcolumn,caption}
\usepackage[noend]{algpseudocode}
\usepackage{algpseudocode}
\usepackage{paralist}
\usepackage{stmaryrd}
\usepackage{tikz}
\usepackage[referable]{threeparttablex}
\usepackage{tabularx}
\usepackage{booktabs}
\usepackage{array}
\usepackage{fancyhdr}
\usepackage{float}
\usepackage{xspace}
\usepackage{pifont}
\usepackage{balance}
\usepackage{xcolor}

\definecolor{codegreen}{rgb}{0,0.6,0}
\definecolor{codegray}{rgb}{0.5,0.5,0.5}
\definecolor{codepurple}{rgb}{0.58,0,0.82}
\definecolor{backcolour}{rgb}{0.95,0.95,0.92}
\lstdefinestyle{mystyle}{
    backgroundcolor=\color{backcolour},   
    commentstyle=\color{codegreen},
    keywordstyle=\color{magenta},
    numberstyle=\tiny\color{codegray},
    stringstyle=\color{codepurple},
    basicstyle=\ttfamily\footnotesize,
    breakatwhitespace=false,         
    breaklines=true,                 
    captionpos=b,                    
    keepspaces=true,                 
    numbers=left,                    
    numbersep=5pt,                  
    showspaces=false,                
    showstringspaces=false,
    showtabs=false,                  
    tabsize=2
}
\newcommand*\circled[1]{\tikz[baseline=(char.base)]{
            \node[shape=circle,fill,inner sep=1pt,scale=0.8] (char) {\textcolor{white}{#1}};}}
\usepackage{listings}
\usepackage{blindtext}
\usepackage{tcolorbox}
\usepackage{graphicx}

\definecolor{codegreen}{rgb}{0,0.6,0}
\definecolor{codegray}{rgb}{0.5,0.5,0.5}
\definecolor{codepurple}{rgb}{0.58,0,0.82}
\definecolor{backcolour}{rgb}{0.95,0.95,0.92}

\usepackage{pifont}
\usepackage{tablefootnote}

\usepackage{graphicx}
\usepackage{adjustbox}
\usepackage[top=2cm,bottom=2cm,left=1.5cm,right=1.5cm]{geometry}

\definecolor{princetonorange}{RGB}{255,143,0}

\definecolor{lightgreen}{RGB}{198, 224, 183}

\definecolor{lightred}{RGB}{240, 205, 176}
\newcommand{\yao}[1]{\textcolor{black}{#1}}

\usepackage{colortbl}
\usepackage{dcolumn}

\definecolor{greenDeep}{RGB}{0,170,0}
\definecolor{greenSlightDeep}{RGB}{0,205,0}
\definecolor{greenShallow}{RGB}{0,255,0}
\definecolor{greenShallower}{RGB}{160,255,0}
\definecolor{orangeShallow}{RGB}{255,190,0}
\definecolor{orangeDeep}{RGB}{255,80,0}
\definecolor{orangeDeeper}{RGB}{255,40,0}
\definecolor{redDeep}{RGB}{255,0,0}

\definecolor{redLight}{RGB}{255,128,114}

\def\zz#1{%
\ifdim#1pt>4.9pt\cellcolor{greenDeep}\else
\ifdim#1pt>3.9pt\cellcolor{greenSlightDeep}\else
\ifdim#1pt>2.9pt\cellcolor{greenShallower}\else
\ifdim#1pt>2.9pt\cellcolor{yellow}\else
\ifdim#1pt>1.9pt\cellcolor{orangeShallow}\else
\ifdim#1pt>1.9pt\cellcolor{orange}\else
\ifdim#1pt>0.9pt\cellcolor{orange}\else
\ifdim#1pt>0.9pt\cellcolor{orangeDeep}\else
\cellcolor{orangeDeep}\fi\fi\fi\fi\fi\fi\fi\fi
#1}

\usepackage{lettrine}

\begin{document}
\IEEEoverridecommandlockouts
\IEEEpubid{\makebox[\columnwidth]{ 979-8-3503-7608-1/24\$31.00 \copyright2024 IEEE \hfill} \hspace{\columnsep}\makebox[\columnwidth]{ }}
\title{\LARGE{RTLCoder: Outperforming GPT-3.5 in Design RTL Generation with Our Open-Source Dataset and Lightweight Solution}\vspace{-.6in}}
\author{\\ \\ \\ \vspace{-.05in}
\IEEEauthorblockN{Shang Liu}\vspace{0.05in}
\IEEEauthorblockA{HKUST,
sliudx@connect.ust.hk}\\ \vspace{-.1in}
\IEEEauthorblockN{Qijun Zhang}\vspace{0.05in}
\IEEEauthorblockA{HKUST,
qzhangcs@connect.ust.hk}
\vspace{-.4in}
\and
\\ \\ \\ \vspace{-.05in}
\IEEEauthorblockN{Wenji Fang}\vspace{0.05in}
\IEEEauthorblockA{HKUST,
wenjifang1@ust.hk}\\ \vspace{-.1in}
\IEEEauthorblockN{Hongce Zhang}\vspace{0.05in}
\IEEEauthorblockA{HKUST (GZ) \& HKUST,
hongcezh@ust.hk}
\vspace{-.4in}
\and
\\ \\ \\ \vspace{-.05in}
\IEEEauthorblockN{Yao Lu}\vspace{0.05in}
\IEEEauthorblockA{HKUST,
yludf@connect.ust.hk}\\ \vspace{-.1in}
\IEEEauthorblockN{Zhiyao Xie$^*$}\vspace{0.05in}
\IEEEauthorblockA{HKUST,
eezhiyao@ust.hk}
\vspace{-.4in}
}


\maketitle

\begingroup\renewcommand\thefootnote{$*$}
\footnotetext{Corresponding Author}
\endgroup


\begin{abstract}

The automatic generation of RTL code (e.g., Verilog) using natural language instructions and large language models (LLMs) has attracted significant research interest recently. However, most existing approaches heavily rely on commercial LLMs such as ChatGPT, while open-source LLMs tailored for this specific design generation task exhibit notably inferior performance. The absence of high-quality open-source solutions restricts the flexibility and data privacy of this emerging technique. In this study, we present a new customized LLM solution with a modest parameter count of only 7B, achieving better performance than GPT-3.5 on all representative benchmarks for RTL code generation. Especially, it outperforms GPT-4 in VerilogEval Machine benchmark. This remarkable balance between accuracy and efficiency is made possible by leveraging our new RTL code dataset and a customized LLM algorithm, both of which have been made fully open-source\footnote{RTLCoder has been open-source in https://github.com/hkust-zhiyao/RTL-Coder. It includes the data generation flow, the complete generated training data set, the model training flow, and the final fine-tuned models (based on both Mistral and DeepSeek).}.

\end{abstract}

\section{Introduction}\label{sec:intro}

In recent years, large language models (LLMs) such as GPT~\cite{gpt4} have demonstrated remarkable performance in natural language processing (NLP). Inspired by this progress, researchers have also started exploring the adoption of LLMs in agile hardware design. Many new LLM-based techniques emerge and attract wide attention in 2023. 
For example, LLM-based solutions are proposed to generate design flow scripts to control EDA tools~\cite{he2023chateda, liu2023chipnemo}, design AI accelerator architectures~\cite{fu2023gpt4aigchip, yan2023viability}, design quantum architectures~\cite{liang2023unleashing}, hardware security assertion generation~\cite{kande2023llm}, fix security bugs~\cite{ahmad2023fixing}, and even directly generate the target design RTL~\cite{chang2023chipgpt, blocklove2023chip, lu2023rtllm, liu2023verilogeval, thakur2023benchmarking, thakur2023autochip, nair2023generating, liu2023chipnemo}.

Among the above explorations, a promising direction that perhaps attracts the most attention is automatically generating design RTL based on natural language instructions~\cite{chang2023chipgpt, blocklove2023chip, lu2023rtllm, liu2023verilogeval, thakur2023benchmarking, thakur2023autochip, nair2023generating, liu2023chipnemo}. Specifically, given design functionality descriptions in natural language, LLM can directly generate corresponding hardware description language (HDL) code such as Verilog, VHDL, and Chisel from scratch. Compared with well-explored \emph{predictive} machine learning (ML)-based solutions in EDA~\cite{rapp2021mlcad}, such \emph{generative} methods benefit the hardware design and optimization process more directly. This LLM-based design generation technique can potentially revolutionize the existing HDL-based VLSI design process, relieving designers from the tedious HDL coding tasks.  

\begin{table}[!t]
      \centering
      \vspace{.1in}
     \hspace{-.1in}
     \setlength{\tabcolsep}{0.5em}
      \renewcommand{\arraystretch}{1.2}
      \resizebox{0.49\textwidth}{!}{
        \begin{tabular}{ |c||c|c|c| } 
        \hline
        \multirow{2}{*}{Works}  &  New Training   &  New LLM  &  Outperform \\
                &   Dataset         &  Model  &  GPT-3.5 \\ 
        \hline
         \hline
         Prompt Engineering &  \multirow{2}{*}{N/A}   &  \multirow{2}{*}{N/A}   & \multirow{2}{*}{N/A}   \\
         \cite{chang2023chipgpt, blocklove2023chip, lu2023rtllm, thakur2023autochip, nair2023generating} &   &  & \\
         \hline
         Thakur et al.~\cite{thakur2023benchmarking} &  \textbf{Open-Source}  &  \textbf{Open-Source} &  No \\
         \hline
         VerilogEval~\cite{liu2023verilogeval}   &  \multirow{4}{*}{Closed-Source}  &  \multirow{4}{*}{Closed-Source} & \multirow{3}{*}{Comparable} \\
         ChipNeMo~\cite{liu2023chipnemo} &   &  &  \\
        ChipGPT-FT ~\cite{chang2024data} & & &\\
        \cline{1-1} \cline{4-4} 
        BetterV~\cite{pei2024betterv} &  &  & {\textbf{Yes}}\\
        \hline
        \hline
        \textbf{RTLCoder}  &  {\textbf{Open-Source}}  &  {\textbf{Open-Source}}  & {\textbf{Yes}}  \\
         \hline
        \end{tabular}
       }
       \vspace{.02in}
        \caption{LLM-based works on automatic design RTL (e.g., Verilog) generation based on natural language instructions.}
               \vspace{-.2in}
        \label{priorWorks}
\end{table}

Table~\ref{priorWorks} summarizes existing work on LLM-based design RTL generation. Some works~\cite{chang2023chipgpt, blocklove2023chip, lu2023rtllm, thakur2023autochip, nair2023generating} focus on prompt engineering methods based on commercial LLMs like GPT, without proposing new datasets or models for RTL code generation. As we will discuss later, reliance on commercial LLM tools limits in-depth research exploration and incurs serious privacy concerns in industrial IC design scenarios. Thakur et al.~\cite{thakur2023benchmarking} generate a large unsupervised training\footnote{Most customized LLM solutions (including RTLCoder) are developed by fine-tuning pre-trained LLMs based on a training dataset about the specific task. In this paper, we use the terms \emph{training} and \emph{fine-tuning} interchangeably.} dataset by collecting Verilog-based projects from online resources like GitHub, then fine-tuning its own model. However, this unsupervised dataset is quite unorganized with a mixture of code and text. Evaluations on a third-party benchmark~\cite{lu2023rtllm} show that the performance of its fine-tuned model is still inferior to commercial tools like GPT-3.5. The VerilogEval~\cite{liu2023verilogeval} from the NVIDIA research team proposes its own labeled training dataset and benchmark, then fine-tunes its own new model. This may be the first non-commercial model that claims comparable performance with GPT-3.5, but according to their authors, neither the training dataset nor fine-tuned LLM model will be released to the public in the near future~\cite{liu2023verilogeval}. Besides these customized RTL-generation solutions, according to our study, all other software code (e.g., Python) generation models like CodeGen2~\cite{nijkamp2023codegen2}, StarCoder~\cite{li2023starcoder}, and Mistral~\cite{jiang2023mistral} are significantly inferior to GPT-3.5 in this RTL generation task. 


Compared with solutions based on closed-source commercial LLM tools like GPT, the open-source LLM solution is vitally important from both research and application perspectives: 1) For research purposes, obviously, closed-source commercial tools prevent most in-depth studies and customizations of this emerging technique. 2) For realistic applications, users of commercial LLM tools unavoidably have data privacy concerns, since all instructions have to be uploaded to LLM providers like OpenAI. In comparison, each user's own local LLM developed based on an open-source solution can eliminate all privacy concerns and also ensure a reliable service. 


\begin{figure*}[!t]
\centering
\includegraphics[width=\textwidth]{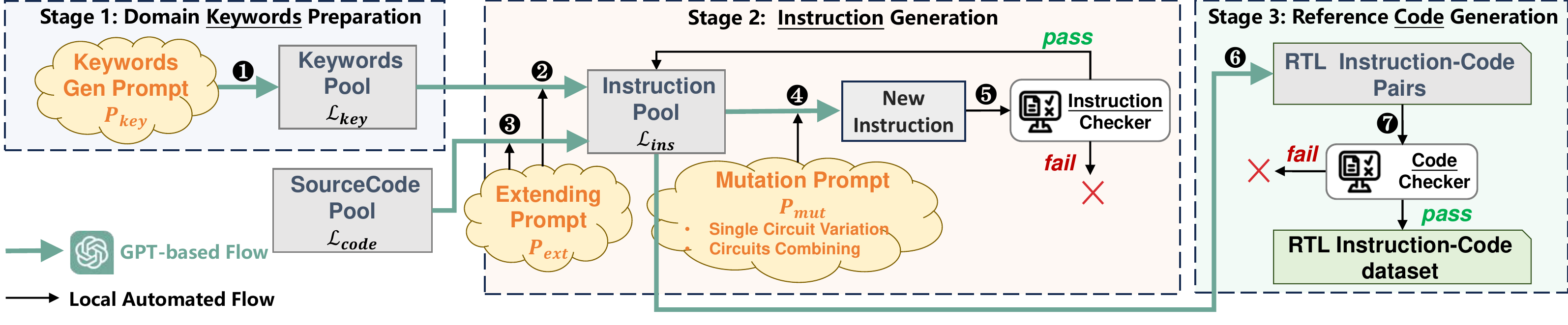}
\vspace{-.25in}
\caption{Our proposed automated training dataset generation flow.} 
\label{fig:flow}
\end{figure*}

However, as mentioned, high-performance open-source RTL generation models are currently unavailable. According to our study, a major challenge is the unavailability of high-quality circuit design data for training: 1) Organized design data is mostly owned by semiconductor companies, who are almost always unwilling to share design data. 2) Design data directly collected online is messy and unorganized, either leading to inferior model performance or requiring prohibitive human efforts to clean the dataset.

In this work, we finally fill this gap with our new open-source LLM solution named \textbf{RTLCoder}. To the best of our knowledge, it is the first open-source LLM that outperforms GPT-3.5 in all representative RTL code generation benchmarks~\cite{liu2023verilogeval, lu2023rtllm}. Our contributions are summarized below. 


\begin{itemize}
    \item Targeting Verilog code generation, we propose an automated flow to generate a large labeled dataset with over 27 thousand diverse Verilog design problems and answers. It addresses the serious data availability challenge in IC design-related tasks, and its potential applications are not limited to LLMs.  LLM directly trained on it can already achieve comparable accuracy to GPT-3.5. 
    \item We introduce a new memory-efficient LLM training scheme based on code quality feedback. It further boosts the ultimate model performance to outperform GPT-3.5, being comparable to GPT-4. Our 7B model can be trained with only four commercial GPU cards. 
 
    \item RTLCoder has been fully open-sourced, including our data generation flow, complete generated dataset, LLM training algorithm, and the fine-tuned model. Considering RTLCoder's lightweight property and low hardware barrier, it allows anyone to easily replicate and further improve based on our existing solution. 
\end{itemize}



\section{Automatic Dateset Generation}
\label{sec:dataset}




In this work, we first propose a new automated training dataset generation flow. Based on this flow, we have generated over 27 thousand training samples, with each sample being a pair of design instruction (i.e., model input) and the reference RTL code (i.e., expected model output). The instruction can be viewed as the input question for LLMs, describing the desired circuit functionality in natural language. The reference code is the expected answer from LLMs, implementing the circuit functionality in Verilog code. We observe that these generated training samples exhibit high diversity and complexity in the RTL-generation domain, encompassing a diverse spectrum of difficulty levels.




We build this automated generation flow by taking full advantage of the powerful general text generation ability of the commercial tool GPT. Please notice that GPT is only used for dataset generation in this work and GPT-3.5 is adopted here. 
The automated dataset generation flow is illustrated in Figure~\ref{fig:flow}, which includes three stages: 1) RTL domain keywords preparation,  2) instruction generation, and 3) reference code generation. We designed several general prompt templates to control GPT generating the desired outputs in each stage.




% ...

\subsection{Stage 1: Keywords Preparation}

The first stage of our data generation flow targets preparing RTL domain keywords for subsequent stages. At process \circled{1} shown in Figure~\ref{fig:flow}, we request GPT to generate keywords related to digital IC design (i.e., commonly used logic components) based on a set of prompts $P_{key}$. We obtain a keyword pool $\mathcal{L}_{key}$ with hundreds of digital design keywords.

Specifically, in this process \circled{1}, to collect a comprehensive range of RTL design task topics, we utilize a tree-like structure with multiple branches to issue queries to GPT. We first prompt GPT at the root node to provide categories and examples of frequently used block keywords in RTL design.
The response from GPT has a tree structure that consists of some related subfields. With the response, we could use the categories and examples as branches to continue prompting GPT for more design keywords within each topic. For example, we can use scripts to ask GPT about more types of the block ``multiplier", it will return more specific design names such as ``Booth multiplier, Wallace tree multiplier, etc.". After this process, we obtain hundreds of keywords related to RTL design in the Keywords pool $\mathcal{L}_{key}$.





\subsection{Stage 2: Instruction Generation}

The second stage targets generating sufficient instructions based on the initial keywords and Verilog source code. At process \circled{2}, we extend existing keywords from $\mathcal{L}_{key}$ to complete instructions. Specifically, we randomly sample one or two keywords from $\mathcal{L}_{key}$ each time, combined with prompts $P_{ext}$, and feed them into GPT to obtain an RTL design instruction.



{In addition to keyword-based instruction generation in process \circled{2}, we also propose to generate instructions based on existing source code collected by us, as shown in process \circled{3}. This is partially inspired by the work of~\cite{wei2023magicoder}. By providing GPT with either part or a complete Verilog code $\mathcal{L}_{code}$ collected by ~\cite{thakur2023benchmarking}, we can inspire it to create a related Verilog design problem. By adopting this new \circled{3} together with \circled{2}, we further enhance the diversity of our dataset by utilizing a vast and varied collection of source code}. 


Process \circled{2} and \circled{3} help generate the initial design instruction pool $\mathcal{L}_{ins}$ based on our customized prompt $P_{ext}$. After generating the initial instruction pool $\mathcal{L}_{ins}$ with hundreds of initial instructions, we will iteratively use mutation methods to significantly augment the scale and complexity of this pool. At \circled{4}, we use $P_{mut}$ to apply two types of mutation operations on instructions sampled from the design instruction library $\mathcal{L}_{ins}$. The process \circled{5} would check every new design instruction using a set of rules and only passed valid instructions are added to $\mathcal{L}_{ins}$. Stage 2 is fully automated and accurate enough to generate a high-quality ultimate instruction pool $\mathcal{L}_{ins}$, including over 50,000 instructions.

In addition, we will further request GPT to generate its reasoning steps (i.e., how it analyzes the generation task step-by-step). These reasoning steps further enhance the detailed information of our instruction pool.

\subsection{Stage 3: Reference Code Generation}




The third stage targets generating the reference code. In the third stage, as shown in \circled{6}, we feed each instruction from $\mathcal{L}_{ins}$ into GPT, generating 5 corresponding reference design codes as the solution candidates.  After that, in \circled{7}, we will evaluate these answers using a code checker. In this work, we adopt an automated syntax checker and only syntax-correct design code can be kept. If all 5 answers fail the syntax checking, this instruction will be discarded. 
Finally, only valid instruction-code pairs are saved as our dataset. Ideally, process \circled{7} should also check whether the functionality of the generated RTL code is consistent with the instruction, but currently generating testbenches for functionality verification cannot be automated. This imperfect automated checking can already filter out the most serious mistakes in the dataset. 



After going through all three stages, we generate the ultimate training dataset with more than 27,000 data samples. An interesting observation is that, although we generate our training dataset based on GPT-3.5, RTLCoder turns out to outperform  GPT-3.5 on representative benchmarks~\cite{lu2023rtllm, liu2023verilogeval}. One important reason is that, for each instruction, we have employed a syntax checker to filter out the obviously incorrect codes generated from GPT-3.5 and retain the largely correct ones for training RTLCoder. This process can be viewed as a refinement of GPT-3.5's Verilog generation capabilities. 

\section{New Training Scheme Incorporating Code Quality Feedback}

 The sequence generation is autoregressive, which means the model always predicts the next token based on its own generated previous ones rather than the reference tokens. Therefore, the traditional model tuning based on maximum likelihood estimation (MLE) would result in a phenomenon named \emph{exposure bias}~\cite{liu2022brio, bengio2015scheduled} and the trained model would still generate many low-quality codes. To alleviate this phenomenon, we propose a new LLM training scheme that incorporates code quality scoring. It further improves the RTLCoder's performance on the RTL generation task.

For each instruction, we will now collect multiple additional code candidates generated by the initial pre-trained model. Then we pack these candidates and the original reference code $y_i$ together as $\textbf{y}_i=\left \{y_{i,k}  \right \}  $, $k=1,2,..,K$, where $K$ represents the number of generated code for one instruction ${x}_i$. Next, all these candidates will be scored by the scoring mechanism $R(x_i, y_{i,k})$ which could be a syntax checker or unit test for functionality check.  We will then obtain a set of score $\textbf{z}_i=\left \{z_{i,k}   \right \}$, $k=1,2,..,K$,  denoting the quality for the code sample $\{y_{i,k}\}$. In the training process, we make the model learn to assign relatively higher generation probabilities to answers with higher scores. 
To further make this training scheme more memory efficient, we decompose the computation graph calculation and use the gradient accumulation-alike method to reduce the space complexity from $O(K)$ to $O(1)$.

\section{Experimental Results} \label{sec:exper}

\subsection{Evaluation Benchmark and Metric}

To evaluate the performance of Verilog code generation, there are two representative benchmarks VerilogEval~\cite{liu2023verilogeval} and RTLLM~\cite{lu2023rtllm}. \yao{The VerilogEval~\cite{liu2023verilogeval} benchmark consists of two parts, EvalMachine and EvalHuman, each including more than 100 RTL design tasks. We follow the original paper~\cite{liu2023verilogeval} and use the widely-adopted $pass@k$ metric.} {The RTLLM V1.1~\cite{lu2023rtllm} benchmark contains 29 RTL design tasks at a larger design scale. We use Synopsys VCS~\cite{vcs} to calculate the scores of the design syntax part and design functionality part separately. \yao{In both parts, following the original benchmark~\cite{lu2023rtllm}, each task is counted as success as long as \emph{any} of 5 trials passes the test. This can be interpreted as pass@5 metric.} In the generation process, we set ${top}_p=0.95$ and $temperature=\left \{  0.2, 0.5, 0.8\right \}$. For all tested models (i.e., baselines, RTLCoder, and ablation studies), we evaluate all 3 $ temperature$ conditions and report the best of each model.

\begin{table*}[!t]
\centering
\hspace{-.2in}
\resizebox{1.02\textwidth}{!}{
\setlength{\tabcolsep}{0.8em}
\renewcommand{\arraystretch}{1.2}
\begin{tabular}{|c|c|c||c|c|c||c|c|c||c|c|} 
\hline
\multirow{4}{*}{Model Type}  & \multirow{4}{*}{Evaluated Model} & \multirow{3}{*}{Num of} & \multicolumn{6}{c||}{VerilogEval Benchmark~\cite{liu2023verilogeval}} &  \multicolumn{2}{c|}{RTLLM V1.1~\cite{lu2023rtllm}}  \\ 
   &   &  \multirow{3}{*}{Params}  & \multicolumn{6}{c||}{(using pass@k metric)} &  \multicolumn{2}{c|}{(using pass@5 metric)}  \\
\cline{4-9}\cline{10-11}
   &     &     & \multicolumn{3}{c||}{Eval-Machine (\%)} & \multicolumn{3}{c||}{Eval-Human (\%)}   &     Syntax-VCS   &   Func   \\ 
\cline{4-9} 
   &     &   &  k=1 &  k=5 &  k=10  &  k=1 &  k=5 &  k=10  &  (\%)   & (\%)   \\
\hline
\multirow{3}{*}{Closed-Source}      & GPT-3.5     & N/A       & 46.7  &69.1  &74.1       & 26.7   &45.8  &51.7                         &89.7    &37.9     \\
\multirow{3}{*}{Baseline}           & GPT4        & N/A       & 60.0  &70.6  &73.5       & \cellcolor{lightgreen}43.5      &\cellcolor{lightgreen}55.8  &\cellcolor{lightgreen}58.9                       &\cellcolor{lightgreen}100    &\cellcolor{lightgreen}65.5      \\
&   ChipNeMo$^\star$~\cite{liu2023chipnemo}  &13B &43.4  &N/A &N/A &22.4  &N/A &N/A   &N/A &N/A              \\
                                    & VerilogEval$^\star$~\cite{liu2023verilogeval}     & 16B     & 46.2   &67.3  &73.7      & 28.8   &45.9  &52.3            & N/A    & N/A        \\
                                    & BetterV$^\star$~\cite{pei2024betterv}  & 7B  &\cellcolor{lightgreen}64.2 &\cellcolor[HTML]{C5D9F1}75.4 &\cellcolor[HTML]{C5D9F1}79.1        & \cellcolor{lightred}40.9 &\cellcolor{lightred}50.0 &\cellcolor{lightred}53.3
                          &N/A &N/A                           \\
\hline
\hline
\multirow{2}{*}{Open-Source}            & Codegen2~\cite{nijkamp2023codegen2}  & 16B  & 5.00  &9.00  &13.9        & 0.90   &4.10  &7.25                          &72.4 &6.90                           \\
\multirow{2}{*}{Baseline}               & Starcoder~\cite{li2023starcoder}     & 15B  & 46.8  &54.5  &59.6      & 18.1    &26.1  &30.4                    &\cellcolor{lightred}93.1 &27.6                          \\
                                    & Thakur et al.~\cite{thakur2023benchmarking}   & 16B    & 44.0  &52.6  &59.2       & 30.3  &43.9  &49.6                 &86.2    &24.1 
                                    \\ 
                                    
\hline
\hline
\multirow{2}{*}{\textbf{Base Model}}                 & Mistral-7B-v0.1~\cite{jiang2023mistral}       & 7B     & 36.9  &48.8  &57.4            &4.49    &12.6  &18.6                           &72.4    &20.7       \\
 &  DeepSeek-Coder-6.7b~\cite{guo2024deepseek}        & 6.7B     & 54.1  &63.8  &67.5            &30.2    &42.2  &46.2                           &89.6    &34.5         \\
 \hline
\hline
\textbf{Less Training Data}
               &  RTLCoder-Mistral-10k          & 7B            &56.5    &66.6  &69.4           &31.7     &42.2  &46.5                            &86.2    &34.5         \\ 
 
 \textbf{(10K Samples)} & RTLCoder-DeepSeek-10k        & 6.7B            & 55.3   &70.4  &76.2            &36.7     &47.0  &50.4                                    &79.3    &37.9   \\
\hline
\hline
\multirow{2}{*}{\textbf{Direct Training}}
               &  RTLCoder-Mistral-Direct          & 7B            & 58.9   &70.0 &74.1            &34.4     &42.3  &45.1                            &89.7    &41.4         \\ 
 
 &  RTLCoder-DeepSeek-Direct        & 6.7B            &59.8    &\cellcolor{lightred}73.6  &\cellcolor{lightred}77.2            &39.1       &48.3                   &51.3              &86.2    &\cellcolor{lightred}44.8   \\
\hline

\multirow{2}{*}{\textbf{RTLCoder}}       

 & \textbf{RTLCoder-Mistral}        & 7B        &\cellcolor[HTML]{C5D9F1}62.5    &72.2  &76.6        & 36.7   &45.5  &49.2                     &\cellcolor[HTML]{C5D9F1}96.6 &\cellcolor[HTML]{C5D9F1}48.3                        \\
                                     & \textbf{RTLCoder-DeepSeek}        & 6.7B        & \cellcolor{lightred}{61.2}   &\cellcolor{lightgreen}76.5  &\cellcolor{lightgreen}81.8         & \cellcolor[HTML]{C5D9F1}{41.6}   &\cellcolor[HTML]{C5D9F1}50.1 &\cellcolor[HTML]{C5D9F1}53.4                     &\cellcolor{lightred}93.1 &\cellcolor[HTML]{C5D9F1}48.3                        \\
\hline
\end{tabular}
}
\begin{tablenotes} \footnotesize
\item$^\star$We cannot directly evaluate VerilogEval~\cite{liu2023verilogeval}, ChipNeMo~\cite{liu2023chipnemo} and BetterV~\cite{pei2024betterv} on RTLLM Benchmark due to closed-source models. We fully understand and respect the authors' privacy concerns. The accuracy values of VerilogEval~\cite{liu2023verilogeval}, ChipNeMo~\cite{liu2023chipnemo}, BetterV~\cite{pei2024betterv}, GPT-3.5, and GPT-4 on the VerilogEval Benchmark~\cite{liu2023verilogeval} are directly cited from the original publication~\cite{liu2023verilogeval, liu2023chipnemo, pei2024betterv}.  
\end{tablenotes} 
\vspace{.03in}
\caption{Performance comparison of RTL code generators on VerilogEval Benchmark~\cite{liu2023verilogeval} and RTLLM Benchmark~\cite{lu2023rtllm}. The top scores ranked 1$^{\text{st}}$, 2$^{\text{nd}}$, and 3$^{\text{rd}}$ in each column are marked in \colorbox{lightgreen}{Green}, \colorbox[HTML]{C5D9F1}{Blue}, and \colorbox{lightred}{Red}, respectively. RTLCoder outperforms GPT-4 on EvalMachine of ~\cite{liu2023verilogeval}. It is only second to GPT-4 on the other benchmarks.}
\label{tbl:result}
\vspace{-.2in}
\end{table*}

\subsection{Model Training}

To ensure a fair evaluation of our proposed RTLCoder, before training, we explicitly examined the similarity between samples in our proposed training dataset and those test cases in benchmarks~\cite{liu2023verilogeval, lu2023rtllm} using Rouge-L metric. Then we get rid of our training samples that are highly similar to test cases during the training process. 

Based on our generated dataset with 27K instruction-code pairs, we choose the latest Mistral-7B-v0.1~\cite{jiang2023mistral} and DeepSeek-Coder-6.7b~\cite{guo2024deepseek} as the basic pre-trained model for finetuning. In all experiments, we opted for the Adam optimizer with $\beta_1$ = 0.9, $\beta_2$ = 0.999, and learning rate $\gamma$ = 1e-5, while abstaining from the use of weight decay. Concurrently, we established a context length of 2048 and a global batch size of 256. We trained the model on only 4 consumer-level RTX 4090 GPUs (24GB each), each of which could only afford $2\times2048$ context length using DeepSpeed stage-2~\cite{rasley2020deepspeed}.

To implement our proposed training scheme, we first generated 3 code candidates for each instruction using a pre-trained model with Beam search method. Then we use Pyverilog~\cite{Takamaeda:2015:ARC:Pyverilog} as the syntax checker to score the code candidates.

\vspace{-.05in}
\subsection{Experiment Results Overview}

\yao{Table~\ref{tbl:result} summarizes the comparison of all relevant RTL generation solutions, including commercial models GPT3.5/GPT4, models customized for Verilog generation~\cite{liu2023verilogeval, thakur2023benchmarking}~\cite{pei2024betterv}, software code generators~\cite{nijkamp2023codegen2, li2023starcoder, jiang2023mistral}, and ablation studies of RTLCoder. }



\yao{In the VerilogEval benchmark~\cite{liu2023verilogeval}, for both EvalMachine and EvalHuman categories, RTLCoder-DeepSeek scores 61.2 and 41.6 respectively. It clearly outperforms GPT-3.5 and is only inferior to GPT-4 among all the models in EvalHuman. Specifically, in the EvalMachine part, RTLCoder-DeepSeek and RTLCoder-Mistral even outperform GPT4 by an absolute value of 1.2\% and 2.5\%. A similar trend can be observed in the RTLLM benchmark V1.1~\cite{lu2023rtllm}. RTLCoder is second only to GPT-4. In summary, RTLCoder outperforms GPT-3.5 and all non-commercial baseline models in most of the metrics. }


\yao{Furthermore, we validate the effectiveness of our proposed dataset and algorithm through an ablation study. The RTLCoder-Mistral-Direct and RTLCoder-DeepSeek-Direct are directly trained with the traditional MLE method. Using our training dataset, they can already significantly outperform the base model and even GPT-3.5 on part of these indexes. Then the RTLCoders trained with our proposed training scheme further outperform those using Direct training method on all benchmarks, indicating that our training method greatly further improves the model performance.}


\yao{We also randomly selected 10K samples from the 27K training dataset to finetune the base models and obtained RTLCoder-Mistral-10k and RTLCoder-DeepSeek-10k respectively. Compared with the two models,  RTLCoders trained on a 27K dataset are clearly superior on all metrics. Increasing the size of the training dataset and enhancing its diversity clearly further improves the model performance.}

\section{Conclusion}\label{sec:concl}

This work presents a fully open-sourced LLM solution named RTLCoder for RTL code generation, achieving state-of-the-art performance in non-commercial solutions and outperforming GPT-3.5. We contribute a new data generation flow and a complete dataset with over 27 thousand labeled samples, addressing the serious data availability problem in hardware-design-related tasks. Also, we contribute a new training scheme based on design quality scoring. It greatly boosts the model performance. RTLCoder's lightweight property and low hardware barrier allow anyone to easily replicate and further improve based on our existing solution.



\section{Acknowledgement}\label{sec:ack}

This work is partially supported by the Hong Kong Research Grants Council (RGC) ECS Grant 26208723.

\bibliographystyle{IEEEtran}
\bibliography{ref}

\end{document}